\documentclass {mn2e}
\usepackage{graphicx}
\title [C-Mira kinematics]{Carbon-Rich Mira Variables: Kinematics and
Absolute Magnitudes}
\author[Feast. Whitelock \& Menzies]{Michael W. Feast$^{1}$\thanks{e-mail:
\tt mwf@artemisia.ast.uct.ac.za}, 
Patricia A. Whitelock$^{1,2,3}$, John W. Menzies$^{2}$\\
$^{1}$ Astronomy Dept. University of Cape Town, Rondebosch, 7701, 
South Africa\\
$^{2}$ South African Astronomical Observatory, P.O. Box 9, Observatory,
7935, South Africa\\
$^{3}$ National Astrophysics and Space Science Programme, University of
Cape Town, Rondebosch 7701, South Africa\\}
\begin{document}
\maketitle
\begin{abstract}
The kinematics of galactic C-Miras are discussed on the basis of the bolometric
magnitudes and radial velocities of Papers I and II of this series.
Differential galactic rotation is used to derive a zero-point for the
bolometric
period-luminosity relation which is in 
satisfactory agreement with that inferred from
the LMC C-Miras. We find for the galactic Miras,
$M_{bol} = -2.54logP + 2.06 (\pm 0.24)$, where the slope is taken from the
LMC. The mean velocity dispersion, together with the data of Nordstr\"{o}m
et al. and the Padova models, leads to a mean age for our sample of C-Miras
of ${1.8 \pm 0.4}$ Gyr and a mean initial mass of
${1.8 \pm 0.2}$ $ \rm M_{\odot}$. Evidence for a variation of velocity
dispersion with period is found, indicating a dependence of period on age
and initial mass, the longer period stars being younger.  We discuss the
relation between the O- and C-Miras and also their relative numbers in
different systems.

\end{abstract}

\section{Introduction}
  This is the third and last of a group of papers dealing with the
properties of carbon-rich Mira variables. In this paper we combine the
photometry and bolometric data of Whitelock et al. (2006, Paper I) and the
radial velocities listed in Menzies et al. (2006, Paper II) to discuss the
kinematics of the C-Miras. A main aim of this paper is to derive estimates
of the absolute magnitudes, ages and initial masses of galactic C-Miras.
These parameters are of importance both for understanding the place of
C-Miras in stellar evolution and for using them as probes of the stellar
populations in other galaxies. Whilst there have been some estimates of the
likely mass range of carbon stars in general (e.g. from studies of their
galactic distribution (Claussen et al. 1987, Groenewegen et al. 1995), and
considerable theoretical work on the production of carbon stars, there has
been little or no observational evidence on the ages and masses of C-Miras
except for the occurrence of three of them in intermediate age clusters in
the Magellanic Clouds (Nishida et al. 2000) and one of somewhat uncertain
period in another LMC cluster (van Loon et al. 2003), but see also the
discussion on IRAS\,04496--6958 in section 6. There has been no evidence as
to whether or not ages and initial masses of C-Miras vary with period. As
regards the luminosities of galactic C-Miras, Whitelock
\& Feast (2000) found that a satisfactory zero-point for the PL($K$)
relation could not be obtained from Hipparcos parallaxes due
mainly to  the small number with significant weight. Other estimates
of C-Mira luminosities were discussed in paper I.

\begin{figure*}
\includegraphics[width=12cm]{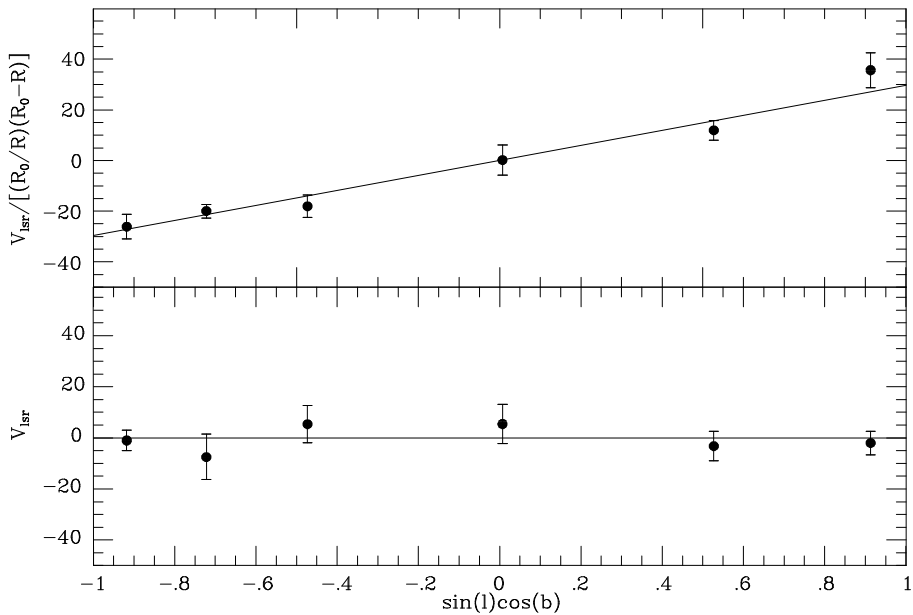}
\caption{ (a). Weighted mean value of
$V_{lsr}/[(R_{\circ}/R)(R_{\circ}-R)]$
plotted against $\sin \ell \cos b$. The line has a slope of
$2A = \rm {29.6\, km\,s^{-1}kpc^{-1}}$
(b). $V_{lsr}$ plotted against $\sin \ell \cos b.$ }
\end{figure*}

\section{Analysis}
The radial velocities used in this analysis are those listed in 
table 4 of Paper II where they have been corrected to our adopted
local standard of rest. The local solar motion adopted was:
$u_{\circ} = +9.3\, \rm {km\,s^{-1}}$ (towards the galactic centre), 
$v_{\circ} = +11.2\, \rm{km\,s^{-1}}$ (in the direction of galactic rotation), 
and 
$w_{\circ} = +7.6\, \rm {km\,s^{-1}}$ (towards the north galactic pole).
These are the values derived by Feast \& Whitelock (1997)
(see also Feast 2000).\\
 
Heliocentric distances were obtained using the period-luminosity relation
derived from LMC observations of C-Miras in Paper I, viz.,
\begin{equation}
M_{bol} = -2.54 \log P + c.
\end{equation}
For an assumed LMC distance modulus of 18.50 the constant term in this
equation was found to be $c = 1.87$. Distances using this value were
tabulated in Paper II (their table 4). One aim of the present analysis is to
estimate the value of $c$ for the galactic sample. The standard general
relation for differential galactic rotation used in the analysis was\\
\begin{eqnarray}
V_{lsr} = 2A(R_{\circ}/R)(R_{\circ} - R) \sin \ell \cos b + \nonumber\\
U_{\circ}\cos \ell \cos b + V_{\circ} \sin \ell \cos b + W_{\circ} \sin b + K
\end{eqnarray}
where: \begin{itemize}
\item $V_{lsr}$ is the radial velocity relative to the local standard of rest
(lsr),
\item $A$ is Oort's constant of differential galactic rotation,
\item $R_{\circ}$ is the distance from the Sun to the Galactic Centre,
\item $R$ is the distance of the Mira from the galactic axis of rotation
(i.e. cylindrical rotation is assumed),
\item $\ell$ and $b$ are the galactic longitude and latitude of the Mira,
\item $U_{\circ}$, $V_{\circ}$ and $W_{\circ}$ are the components of any
group motion with respect to the adopted standard of rest and in the same
directions as the lsr components given above.
\item The $K$ term tests for possible systematic effects in the radial velocities.
\end{itemize}

We adopt $R_{\circ} = 7.62 \pm 0.32\, \rm{kpc}$ from Eisenhauer et al. (2005). 
This is the
value derived from the motions of objects close to the central black hole.
We also adopt $A = 14.82 \pm 0.84\, \rm{km\,s^{-1}\,kpc^{-1}}$ from the
analysis of the Hipparcos proper motions of galactic Cepheids 
(Feast \& Whitelock 1997). We carried out some preliminary calculations
including in equation 2  some higher order rotation terms, 
$A_{2}$ and $A_{3}$
in the nomenclature of Pont et al. (1994), and using the values obtained 
for them by these workers
from an analysis of Cepheid radial velocities. However, their values were
not confirmed in the analysis of Cepheid proper motions (Feast \&
Whitelock 1997) and may be due to group motions on the kiloparsec scale
which are known to occur for young objects. Group motions 
of Cepheids do not necessarily
apply to an older  population such as the C-Miras. Furthermore, since we 
restrict
our main discussion to objects having $|R - R_{\circ}| \leq 2\, \rm {kpc}$,
the effect of the higher order terms is minor.

That differential galactic rotation is clearly present in our sample
can be seen from Fig.~1. Here we plot,
the mean values of the 
quantity,
$V_{lsr}/[(R_{\circ}/R)(R_{\circ} - R)]$,
weighted by $[(R_{\circ}/R)(R_{\circ} - R)]^{2}$, against $\sin \ell \cos b$
for the stars with $|R - R_{\circ}| \leq 2$ kpc.
For distances we have used equation 1 with $c=1.87$ from the LMC. 
The points have been grouped by $\sin \ell \cos b$ so
as to obtain approximately equal weights for each point. The standard
errors are shown. For differential galactic rotation we expect a line
of slope $2A$ as is indeed seen. Also plotted 
are grouped values of  $V_{lsr}$ against
$\sin\ell\cos b$ which shows that there is no effect of
asymmetric drift (i.e. $V_{o}\sim 0$ as shown below). Note that in this
and subsequent work we omit the high velocity C-Mira V CrB which is
discussed in section 5.

\begin{table*}
\centering
\caption{Preliminary Solutions}
\begin{tabular}{rrrrrrrr}
\hline
$c\;\;$ & $A\;\;\;\;\;\;\;\;$ & $U_{o}\;\;\;\;$ & $V_{o}\;\;\;\;$ 
& $W_{o}\;\;\;\;$ & $K\;\;\;\;$ & N$\;\;$ & $\log P$\\
mag & $\rm km\,s^{-1}\,kpc^{-1} $ & $\rm km\,s^{-1}$ & $\rm km\,s^{-1}$ &
$\rm km\,s^{-1}$ & $\rm km\,s^{-1}$ & &\\
1.87 & $13.4 \pm 1.4$ & $-1.0 \pm 3.0$ & $+1.5 \pm 2.4$ & $-3.5 \pm 7.1$ &
$-0.7 \pm 1.8$ & 146 & 2.718 \\
1.87 & $13.4 \pm 1.4$ & $-1.1 \pm 3.0$ & $+1.7 \pm 2.4$ & $-3.6 \pm 7.0$ &
0 & 146 & 2.718 \\
2.06 & $14.8 \pm 1.5$ & $-0.7 \pm 3.0$ & $+0.7 \pm 2.4$ & $-3.8 \pm 7.0$ &
$-0.2 \pm 1.8$ & 149 & 2.717\\
2.06 & $14.8 \pm 1.5$ & $-0.7 \pm 3.0$ & $+0.8 \pm 2.3$ & $-3.8 \pm 7.0$ &
0 & 149 & 2.717 \\
\hline
\end{tabular}
\end{table*}

Table 1 gives the results of solving equation 2, with distances based
on equation 1 
and $c=1.87$ , for the 146 stars with $|R_{\circ} - R| < 2 \rm kpc$
both with $K$ left free and with it put to zero. The table shows that
within the uncertainties $U_{o}, V_{o}, W_{o}$ and $K$ are zero and this
has been assumed in the following. Also shown 
for comparison in Table 1 are the results
using distances based on equation 1 but with the constant term 
($c$) taken as 2.06, the value we derive later. Individual distances
with this value of $c$ are also listed in table 4 of Paper II. 
In Table 1, $\log P$ is the 
weighted mean period 
\footnote{The different numbers of stars in the various solutions
of tables 1 and 2 are due to the movement of stars into or out of the
$|R - R_{\circ}|$ annulus as a result of the changes in the adopted
distance scale.}.
 
With $U_{\circ}$, $V_{\circ}$, $W_{\circ}$ and $K$ put equal to zero
we have solved equation 2 for $A$ using a range of values of the
constant term on the right hand side of equation 1. 
Results are shown in Table 2. For each change in the PL
zero-point the interstellar absorption for each star has been iterated
as described in Paper I. The number of stars in each solution is given as
well as the weighted mean log period. Adopting 
$A = 14.82 \pm 0.84\, \rm km\,s^{-1}kpc^{-1}$ from Cepheids 
(Feast \& Whitelock 1997)
one finds for the PL zero-point, 
$$c = 2.06\pm 0.24.$$
The estimated error includes both the uncertainties in the values of
$A$ from the C-Miras and in that from Cepheids. A solution adopting
$R_{\circ} = 8.0 \,\rm kpc$ gives $c= 2.07$ indicating that uncertainty
in $R_{\circ}$ does not significantly affect our result.

The data were subdivided by period to see whether it was possible to get
an independent slope for the PL relation. However, the results were
inconclusive.

The 149 stars used to derive the constant term 2.06 have a weighted mean 
$\log P = 2.717$.
Fig.~2 shows a plot of the LMC C-Mira data 
discussed in Paper I with an assumed LMC modulus
of 18.50 together with  a point and its error bar corresponding to the
galactic result.
It can be seen that
within the uncertainties the derived absolute magnitude does not
differ significantly from that of stars of the same $\log P$
in the LMC at its adopted distance.
In principle a bias correction is necessary to the galactic zero-point
calculated above (see for instance Feast 2002). However, provided
the true scatter of absolute bolometric magnitudes about
the period-luminosity relation is as small in the Galaxy as in the LMC
($\sigma = 0.17$ mag, see Paper I), this correction is negligibly small.
Numerous recent discussions (e.g. Feast 2003b) suggest that the 
uncertainty in the distance modulus of the LMC is 0.1mag or less.
This is better than that of our galactic determination. Thus, since the two
PL zero-point determinations agree within the errors we decided
it was best to use the LMC calibration for the discussion of paper I.
For convenience distances based on both zero-points are given for
the radial velocity sample in paper II. In the present paper the
galactic calibration is used in the determination of the dispersions
so as to be consistent in the kinematic analysis.

\begin{table}
\centering
\caption{PL Zero-Point Determination}
\begin{tabular}{cccc}
\hline
$c$ & $A$ & N & $\log P$\\
mag & $\rm km\,s^{-1} kpc^{-1}$ &  &  \\
1.8 & $13.2 \pm 1.3$ & 144 & 2.714 \\  
1.9 & $13.6 \pm 1.4$ & 146 & 2.718 \\
2.0 & $14.4 \pm 1.4$ & 148 & 2.717 \\
2.06 & $14.8 \pm 1.4$ & 149 & 2.717 \\ 
2.1 & $15.1 \pm 1.4$ & 149 & 2.717 \\
2.2 & $15.3 \pm 1.5$ & 151 & 2.716 \\
\hline
\end{tabular}
\end{table}

\section{Velocity Dispersions, Ages and Masses}
Using equation 1 with c= 2.06 
and equation 2 with $A = 14.82$ 
and $U_{o}, V_{o}, W_{o}$ and $K$ put equal to zero,
we calculated the
residual radial velocities for the 149 stars (excluding 
the high velocity star V CrB) with
$|R - R_{\circ}| \leq 2.0\, \rm {kpc}$. These were then analyzed for the
velocity dispersions $\alpha$, towards the galactic centre,
$\beta$, at right angles to this and in the galactic plane, and
$\gamma$ perpendicular to the galactic plane. Since the stars involved 
are primarily at low galactic latitude the value of $\gamma$ is effectively
indeterminate from our data. 
Thus in group 2, 30 percent of the weight of $\gamma$ is
in one star which happens to have a moderately large velocity
residual making both the derived $\gamma$ and its standard error
uncertain. In group 1, the square of $\gamma$ (the quantity actually
determined) is negative, though not significantly so.
We therefore adopt 
$\gamma = 12 \, \rm {km\,s^{-1}}$, a value appropriate for the
values of $\alpha$ and $\beta$ found (e.g. Nordstr\"{o}m et al. 2004).
The value adopted will not affect our
results in any significant way. The dispersions were calculated for all the 
stars in the annulus chosen and for the same stars divided into three
groups by period 
(Group 1, $P < 470$ days, Group 2, $470< P< 570$ days, Group 3 $P > 570$ days). 
Table 3 contains the results. It gives the values of 
$\alpha$ and $\beta$ and the value of $\alpha$ derived on the assumption that
$\beta = 0.63 \alpha$. This is the ratio found in the extensive discussion
of local F and G type dwarfs by Nordstr\"{o}m et al. (2004). We use these
latter values of $\alpha$ in the following. The discussion of the 
uncertainties in the adopted radial velocities in Paper II indicates that
any contribution of observational scatter to the calculated dispersions
will be small. Any such correction would slightly decrease the ages we
calculate below.

\begin{table}
\centering
\caption{Velocity Dispersions}
\begin{tabular}{cccccc}
\hline

Group & All & 1 & 2 & 3 & \\
N & 149 & 49 & 50 & 50 &\\
$\log P$ & 2.717 & 2.572 & 2.717 & 2.816 &\\
\hline
$\alpha$ & $26 \pm 3$ & $35 \pm 4$ & $21 \pm 5$ & $22 \pm 3$ &
$\rm km\,s^{-1}$\\
$\beta$  & $18 \pm 3$ & $19 \pm 6$ & $17 \pm 5$ & $17 \pm 3$ &
$\rm km\,s^{-1}$\\
$\gamma$ & $17 \pm 12$ & $ -$       & $50 \pm 11$ & $23 \pm 11$ &
$\rm km\,s^{-1}$\\
\hline
$\alpha$ & $26 \pm 3$ & $34 \pm 4$ & $22 \pm 5$ & $22 \pm 3$ & 
$\rm km\,s^{-1}$ \\
$\beta$  & $18 \pm 3$ & $18 \pm 6$ & $19 \pm 4$ & $17 \pm 3$ &
$\rm km\,s^{-1}$\\
$\gamma$ &  12 (fixed) & & & & $\rm km\,s^{-1}$\\
\hline
$\alpha$ & $27 \pm 2$ & $32 \pm 3$ & $24 \pm 3$ & $24 \pm 2$ & 
$\rm km\,s^{-1}$\\
$\beta$ & $\beta = 0.63 \alpha$ & & & & $\rm km\,s^{-1}$\\
$\gamma$ & 12 (fixed) & & & & $\rm km\,s^{-1}$\\

\hline
\end{tabular}
\end{table}

Nordstr\"{o}m et al. (2004) (their fig. 31) show the relation between
their $\log \sigma_{U}$ (equivalent to $\log \alpha$) and $\log \rm {Age}$.
The numerical values used for this figure were kindly provided to us
by Dr Holmberg. Fitting a straight line to these (log - log) data we 
estimate the ages of the various groups in Table 3 together with their
uncertainties. Using the tabulation of the Padova models (Girardi et al. 2000)
which were also used by Nordstr\"{o}m et al., we then estimate the mean 
initial masses
of the various groups assuming solar metallicity. 
These results are given in Table 4. 
The errors correspond to the standard errors of $\alpha$ and
do not take into account other sources of uncertainty (e.g. of the
Padova models).

The ages derived are similar to those derived by Nishida et al. (2000)
for C-Miras in Magellanic Cloud clusters. They adopt
ages of $1.6\pm 0.4, 1.7\pm 1.0$ and $2.0 \pm 0.2$ Gyr for their three clusters
in which the Miras have periods of 526, 450 and 491 days.
Note that the main-sequence turn-off (TO) masses they adopt 
($\rm 1.6M_{\odot}$) are 
similar to the ones we would derive, e.g. the Girardi et al. 
TO mass is $\rm 1.4\,M_{\odot}$
for solar metallicity stars of age 2 Gyr (the initial masses of stars
at the AGB tip at this age is larger, as shown in Table 4).
van Loon et al. (2003) classify LI-LMC 1813 
in the cluster KMHK 1603 as a C-Mira with
a period of $\sim$680 days and suggest an
age of 0.9-1.0 Gyr and $\rm M_{ZAMS}$ of 2.2 $\rm M_{\odot}$ for it.
This is consistent with our results.
Somewhat smaller initial masses ($\rm 1.3\, M_{\odot}$) 
for dust-enshrouded AGB variables (both oxygen- and carbon-rich)
were estimated 
by Olivier et al. (2001) from scale height considerations.

\begin{table}
\centering
\caption{Estimated Ages and Masses}
\begin{tabular}{ccccc}
Group & $\log P$ & $\alpha$ & Age & $\rm M_{init}/M_{\odot}$ \\
      &          & $\rm km\,s^{-1}$ & Gyr & \\
\hline
All &  2.717 & $27 \pm 2$ & $1.8 \pm 0.4$ & $1.8 \pm 0.2$ \\
1   &  2.572 & $32 \pm 3$ & $3.1 \pm 0.9$ & $1.5 \pm 0.2$ \\
2   &  2.717 & $24 \pm 3$ & $1.3 \pm 0.5$ & $2.1 \pm 0.3$ \\
3   &  2.816 & $24 \pm 2$ & $1.3 \pm 0.3$  & $2.1 \pm 0.2$ \\
\hline
    
\end{tabular}
\end{table}

\begin{figure}
\includegraphics[width=8.5cm]{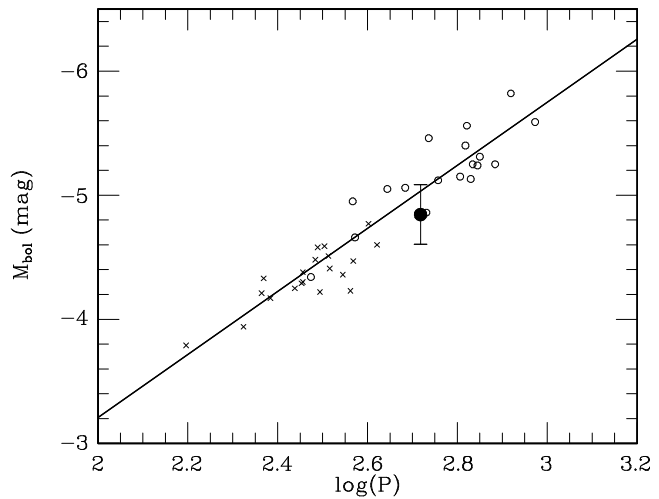}
\caption{The $M_{bol}$ versus $\log P$ relation for C-Miras. Crosses
and open circles are for LMC stars (modulus 18.50) (see Paper I). The filled
circle and error bar are for the present Galactic result.}
\end{figure}

A new set of
isochrones has been published by Pietrinferni et al. (2004). These differ
significantly from those of Girardi et al. (2000) in the age range 
of concern for us
(see Pietrinferni et al. section 6 and fig. 8).
A Girardi solar-composition isochrone for 1.8 Gyr is best fitted with a
Pietrinferni isochrone of age 2.5 Gyr. 
The TO masses are about 0.4 $\rm M_{\odot}$ greater in the
Pietrinferni et al. models.
The difference between the two
sets of isochrones is attributed by Pietrinferni et al. partly to
differences in model input physics and partly to the interplay of
the overshooting treatment with the input physics. It is clear from this
that the absolute ages of our stars (and of Magellanic Cloud clusters
with C-Miras) is still uncertain at the $\sim 0.5$ Gyr level. 

The results in Table 4 show that the shortest period stars (Group 1)
are older and of smaller initial mass than the longest period ones.
Groups 2 and 3 appear to have the same mean ages and masses. However,
given the uncertainties one cannot rule out the possibility that a
trend of age with period continues through the whole sequence
of periods. In any case the results indicate that in the mean the
stars of groups 2 and 3 are not the evolutionary products of
group 1 stars. Similar conclusions apply to the O-Miras
(see e.g. Feast \& Whitelock 2000a).
The relation between the O- and C-Miras is discussed in section 4.

The above discussion is based on the smooth variation of velocity 
dispersion with age which Nordstr\"{o}m et al. (2004) deduce from their data.
However, Quillen \& Garnett (2000,
2001) derived the motions of stars in the Edvardsson et al. (1993)
metallicity study of local F and G dwarfs 
and suggested that the velocity dispersions were constant
in the age range 3 to 9 Gyr (see also Freeman \& Bland-Hawthorn 2002).
If this is so the age of our group 1 (the shortest period group) could be near
3 Gyr 
but could also  be much older. In the latter case, there would
be a rapid change of age with period.
A comparison of the mean data points used by Quillen \& Garnett
with those used by Nordstr\"{o}m et al. shows that there is no very
strong disagreement between them. However, 
the relatively smooth variation
of asymmetric drift and velocity dispersions with period for the O-Miras
(Feast \& Whitelock 2000a) may be difficult to understand if one
adopts the conclusions of Quillen \& Garnett
(see section 4). Also the results of
Binney et al. (2000) on the motions of main sequence and subgiant stars from
the Hipparcos catalogue are consistent with a smooth variation of
velocity dispersion with age such as that found by Nordstr\"{o}m et al.
It should, however, be stressed that whether one adopts the conclusions of 
Nordstr\"{o}m et al.
or those of Quillen \& Garnett, our results suggest a dependence of 
C-Mira period on age at least at the short period end of the
sequence.

\section{The O- and C-Miras and Metallicity}
In this section we discuss the relationship of the O-Miras 
to the C-Miras.

The ratio of
carbon stars to late type oxygen-rich stars (M types)
has been used by a number of workers as a measure 
of the mean 
metallicity
of a galaxy, carbon stars being more frequent in low metallicity
systems. A calibration of this relation has recently been obtained
by Battinelli \& Demers (2005) from local group galaxies.

However, at least in the case of Mira variables,
this ratio must also be affected by the age distribution
of the populations studied. 
Both the asymmetric drift and the velocity 
dispersion of O-Miras are a function of period (e.g. Feast 1963,
Feast \& Whitelock 2000a). Table 5 shows the velocity dispersion, $\alpha$,
for O-Miras as a function of period using the data on their space motions
given by Feast \& Whitelock. To be compatible with the C-Mira results
we have taken a mean of $\Sigma_{V_{R}} (\equiv \alpha)$ and
$1.59 \Sigma_{V_{\theta}} (\equiv \beta/0.63)$. 
Taken together with the occurrence of short period Miras in 
globular clusters
(e.g. Feast et al. 2002) this indicates ages ranging from that of
metal-rich globular clusters at the short period end to $\sim 2$ Gyr
at $\sim 450$ days and probably younger for the longer period (OH/IR)
Miras. The last column of Table 5 shows the number of stars in each
of the period groups. A comparison with Table 4 shows that  
the optically selected sample
of O-Miras discussed by Feast \& Whitelock are, at least in the main,
a much hotter and older population than the C-Miras discussed above
\footnote {It is worth noting that if we adopt the velocity dispersion - age
relation favoured by Quillen \& Garnett (2000, 2001) 
the kinematics of the 
bulk of the 
optically selected local O-Miras would fall in the discontinuity
in their relation, between their thin and thick disc components.
There would thus be no local dwarf stars 
with similar kinematics as counterparts to these
O-Miras, a surprising result.
This problem is not evident if the Nordstr\"{o}m et al. relation
is adopted.}.

\begin{table}
\centering
\caption {The Velocity Dispersion of O-Miras}
\begin{tabular}{ccccc}
\hline
FW Group & $\alpha$ & P & log P & N\\
         & $\rm km\,s^{-1}$ & days & \\
\hline
2 & $76 \pm 8$ & 175 & 2.243 & 17 \\
3 & $69 \pm 6$ & 228 & 2.358 & 24 \\
4 & $46 \pm 4$ & 272 & 2.435 & 26 \\
5 & $50 \pm 4$ & 324 & 2.511 & 40 \\
6 & $45 \pm 4$ & 383 & 2.583 & 32 \\
7 & $28 \pm 4$ & 453 & 2.656 & 15 \\
\hline
\end{tabular}
\end{table}  

As an example of the effect of age on the ratio of O- to C-Miras one
can consider the solar neighbourhood.
In an approximately distance-limited sample 
of optically selected Miras
in the solar neighbourhood
Wood \& Cahn (1977) estimated there were about 14 times more O-Miras
than C-Miras. Optical samples are weighted to the shorter period (older)
stars. On the other hand, in the case of the longer period (younger) 
dust enshrouded
stars a distance limited sample shows roughly equal numbers of
O- and C-Miras (e.g. Olivier et al. 2001, fig. 16) 
\footnote{At the very long
period end of the distribution, $P > 1000$ days, there are 
(OH/IR) O-Miras both in the solar neighbourhood (Olivier et al. 2001 fig. 3)
and also in the LMC (Whitelock et al. 2000) but few, if any, C-Miras.
IRAS 01144+6658 is a possible exception (See section 6).}.
Clearly the ratio of O- to C-Miras in a given population must
be at least partly dependent on age. Since the younger stars
are unlikely to be metal-weak compared with  the older ones (note the
period-metallicity relation for O-Miras (e.g Feast \& Whitelock 2000b),
the above result is not consistent with a simple 
decrease in the O-Mira to C-Mira ratio with decreasing metallicity.

The above two paragraphs indicate that the ratio of C-Miras to O-Miras
in a population will depend on the age distribution in that population.
That other factors can affect this ratio is shown by the galactic Bulge.
The Bulge contains many M-type giants and O-Miras with a wide range of 
periods, up to 700 days in the SgrI window (Glass et al. 1995) and
even longer near the galactic centre (Wood et al. 1998). In contrast there
is an absence of C-Miras and normal carbon stars in general from 
the Bulge 
\footnote{The C-Mira IRAS 17463--4007 with {$\ell = 350^{\circ}.8$},
{$b=-6^{\circ}.5$} and {$d = 10.6$} kpc (see paper II) might lie within 
or near the Bulge.}.
In the past this was frequently assumed to be a metallicity effect
(e.g. King 1993) with the Bulge being super-metal rich compared to the
solar neighbourhood. However, it is now known that the Bulge and
the local disc population have rather similar metallicities.
Rich \& Origlia (2005) find a mean $[Fe/H] = -0.19$ for Bulge
M-type giants and Fullbright et al. (2005) obtained a
mean $[Fe/H] = -0.10$ for Bulge K giants
(see also McWilliam \& Rich 2003). These values have to be 
compared with a mean thin disc metallicity of $[Fe/H] = -0.14$
(Nordstr\"{o}m et al. 2004). There have in fact been suggestions
(Haywood 2001, 2002, Taylor \& Croxall 2005) that the mean metallicity
of the local thin disc could be even greater than this 
($[Fe/H] \sim -0.04$).
  
Our local sample of C-Miras has (Table 4) the same kinematics
as the longer period local O-Miras (e.g. Group 7 of Table 5). These
local  subgroups of O- and C-Miras are thus presumably of the same age.
The presence
in the Bulge of O-Miras in the Group 7 period range and the absence of C-Miras
cannot therefore be ascribed to age difference nor, in view of the above
discussion to different overall metallicities of their parent populations
\footnote{As has long been known, the range of O-Mira periods in
the Bulge implies a range of ages, despite the conclusion of
Zoccali et al. (2003) from colour-magnitude
diagrams that most of the Bulge is very old.}.

However, there is a difference between the Bulge and the local populations
which seems to explain this situation. It was found by Glass et al. (1995)
that the $JHK$ colours of Bulge O-Miras were different from those in the
LMC. This was explained (Feast 1996, Feast \& Whitelock 2000b) as due to
stronger $\rm H_{2}O$ bands at a given period in the Bulge stars. A rough
estimate showed that this could be accounted for by a difference in oxygen
abundance between the two parent populations of a factor of about 2.5 (0.4
dex). It is now known from spectroscopic work that oxygen (an $\alpha$
element) is enhanced in Bulge stars. Rich \& Origlia (2005) find [O/Fe]
$\sim +0.3$ in Bulge M-giants (see their fig. 5).  On the other hand [O/Fe]
tends to be lower in the LMC than in the solar neighbourhood except for very
metal-poor objects (Hill 2004, Cole et al. 2005). It therefore seems a
reasonable assumption that this oxygen overabundance in the Bulge is
sufficient to keep O/C $> 1.0$ through the dredge-up phases of Bulge giants
and thus to prevent the occurrence of carbon stars. It is worth noting that the
evolutionary tracks of Salasnich et al. (2000) suggest that at the
intermediate ages of the C-Miras and the longer period O-Miras, an $\alpha$
element enhancement will not significantly affect the age/mass relation.

In the light of the discussion in this section we can expect the ratio
of O- to C-Miras (and probably the relative frequency of carbon to oxygen
rich giants generally) in different systems to be
a function of overall metallicity,
age and detailed abundance ratios such as [O/Fe]. 
Thus the correlation between overall metallicity
and the ratio of carbon-rich to oxygen-rich giants in local
group galaxies (e.g. Battinelli \& Demers 2005) is
of particular interest since it would seem to imply
other underlying correlations (i.e. between age, overall metallicity
and detailed abundance ratios).

The galactic C-Miras belong kinematically to the thin disc and the bulk of 
the thin disc stars, including those of the relevant ages, have only
slight metal deficiencies. 
As just noted, Nordstr\"{o}m et al. (2004) estimate
an overall mean value of [Fe/H] = --0.14 with a dispersion of 0.19.
If the production of C-Miras were strongly dependent on metallicity,
this
would skew the mean abundance of these stars to lower than that of the
general thin disc. In fact Abia et al. (2001) give spectroscopic
estimates near solar for two of them (V Oph with [M/H] = 0.0 and
VX Gem with [M/H] = --0.15) and for Galactic carbon stars in general
they find near solar metallicities. 

Abundance determinations for C-Miras
are difficult but a
spectroscopic study of a large sample might show whether
there is any bias in their metallicity distribution compared with old
disc stars of the same age.   

\section{Stars with high velocities}
It has been known for many years that the C-Mira V CrB has a high velocity
perpendicular to the galactic plane (e.g. Eggen 1969). 
Using the radial velocity in Paper II table 4, the Hipparcos proper motion
components and a PL distance, the velocity components relative
to the local standard of rest are $u = +28$, $v = -67$ and 
$w = -92 \,\rm km\,s^{-1}$. 
It also seems to 
be underabundant in metals. Abia et al. (2001) give [M/H] = --1.35.
Kipper (1998) gave [Fe/H] = --2.12. The large difference is no doubt
partly due to the problems of spectroscopic analysis of such a cool star.
Nevertheless, it seems safe to assume that it has an unusually low
metallicity for a galactic C-Mira. It seems most likely to be an extragalactic
interloper. 
There are C stars in the region of the galactic halo which are likely to
have an extragalactic origin (e.g. Ibata et al. 2001) and at least
in the Sgr dwarf galaxy some of these are known to be Miras 
(Whitelock et al. 1999).
 
\begin{table*}
\centering
\caption{Stars with High Residual Velocities}
\begin{tabular}{rrrrcc}
\hline
Star & \multicolumn{1}{c}{$\ell$} & \multicolumn{1}{c}{$b$} & $V_{lsr}$ & 
\multicolumn{1}{c}{Residual} & \multicolumn{1}{c}{P} \\
     & \multicolumn{2}{c}{(deg)} & \multicolumn{2}{c}{($\rm km\,s^{-1}$)}  &
(day)\\
\hline
$|R_{o} - R| < 2.0 \rm kpc$ & & & & &\\
\hline
V CrB & 63.27 & +51.23 & --102 & --96 & 358\\
ZZ Gem & 187.60 & +5.54 & +67  & +62 & 316\\
KY Cam & 140.93 & +3.47 & --88  & --61 & 477\\
AZ Aur & 172.30 & +8.16 & +71  & +76 & 416\\
\hline
$|R_{o} - R| > 2.0 \rm kpc$ & & & & & \\
\hline
IRAS 06088+1909 & 191.47 & +0.27 & +64 & +54 & 493 \\
RT Gem          & 195.73 & +7.33 & +101 & +84 & 350 \\
IRAS 16171--4759 & 334.86 & +1.35 & +40 & +78  & 560 \\
IRAS 17222--2328 & 2.13   & +6.76 & --61 & --65  & 603 \\
\hline
\end{tabular}
\end{table*}

V CrB is the only star which we have omitted from the discussions of the
earlier sections. However, for completeness we list in Table 6 the stars
which have residuals from the galactic rotation solutions of greater than
$|50|\, \rm {km\,s^{-1}}$. The table is in two sections with stars in the
$|R_{o} -R| < 2.0 \,\rm kpc$ annulus separated from those outside.
In view of the velocity dispersions involved it is not surprising that
there are a few stars with relatively large residuals, and outside the
annulus there may well be significant deviations from the model adopted. 
It is however somewhat intriguing that in the table there are four
stars in the general direction of the  galactic anticentre and one in the
general direction of the galactic centre where the effects of the model
are less significant. These stars 
(ZZ Gem, AZ Aur, IRAS06088+1909, RT Gem, IRAS 17222--2328) 
all have a large radial component outwards from the galactic centre.
It would be interesting to
study their space motions and to see whether this is
anything more than a statistical chance effect. A somewhat more complex outward
motion was found for short period O-Miras (Feast \& Whitelock 2000a,
Feast 2003a) which was attributed to a bar-like motion. It should however
be noted that the kinematics show that the short period O-Miras and the
C-Miras discussed in this paper belong to quite different populations.

\section{Discussion}
   In section 2 we found that if we assume the C-Miras follow a period,
absolute bolometric magnitude relation of the same slope as for the C-Miras
in the LMC, then the zero-point of this relation is, within the
uncertainties, the same in the Galaxy as in the LMC
despite possible differences in metallicities. This is a useful result
since it suggests that C-Miras may be used with some confidence as
distance indicators in populations of different metallicity. This
is of importance for extragalactic applications. It may well be that
in some applications (e.g. extragalactic work), it will be necessary
to derive the apparent bolometric magnitudes from near infrared observations
alone. In that case the bolometric corrections derived in Paper I should
be valuable.

Our kinematic results indicate that the C-Miras, at least in the period
ranges we cover, come from a rather restricted range of ages and initial
masses. The low mean initial masses and intermediate ages we find  
are in agreement with 
those estimated for the Magellanic Cloud clusters containing
C-Miras. 
They are also in agreement with the results of Kahane et al. (2000).
These authors compared isotope abundances in the C-Mira CW Leo
(IRC+10216) with theoretical predictions and found that for solar
metallicities their $5\, M_{\odot}$ model disagreed with the observations
but that models with masses less than about $3\, M_{\odot}$ were
satisfactory.

Considerably higher initial masses have
sometimes been suggested at least for the longer period C-Miras.
Bergeat et al. (2002) suggest $\sim 4\, M_{\odot}$, 
whilst Raimondo
et al. (2005) suggest that the bulk of the long period variables
(in which they include non-Miras)
in the SMC are of age $\sim 0.3$ to $0.5$ Gyr, 
much younger than our estimated ages for C-Miras.
Raimondo et al. suggest that the SMC variables
were formed in a burst of
star formation believed to have taken place  
at about that time (Harris \& Zaritsky 2004).
Our mean age, $\sim 2 \,\rm Gyr$ (or the somewhat larger value
suggested by the Pietrinferni et al. models), if it applies to the
SMC C-Miras, would  not be consistent with them being formed
in such a burst. They would in fact be nearer in age to
the burst that Harris \& Zaritsky date at 2.5 Gyr ago.

It should be noted that 
whilst the $K - \log P$ diagram of Raimondo et al.
(2005)(their fig. 8) shows the Wood ``C" sequence (on which the C-Miras
lie) in broad agreement with the PL($K$) relation of Feast et al. (1989)
shifted to the SMC distance, though much broader due to their dependence
on single near IR observations,
their $M_{bol} -\log P$  diagram
(their fig. 15) has a Wood ``C" sequence which is much steeper,
and displaced from our PL($M_{bol}$)
relation. 
This may be connected with the bolometric corrections for carbon
stars that they adopt from Bergeat et al. (2001, 2002).

Our results, considered together with those
on the C-Miras in Magellanic Cloud clusters, suggest that the C-Miras should 
provide a  mean age for a population in which they occur
(say in an extragalactic system). In addition the suggestion in
our data that initial mass and age are a function of the period of these
stars, 
at least at the shorter periods,
indicates that a comparison of the distribution of C-Mira periods
in different stellar systems will enable a comparative study of star formation
in the interval $\sim 1.5$ to $\sim 3.0$ Gyr ago.

{\bf The analysis of radial velocities and photometry in this paper is based
on
the assumption that the bulk of galactic carbon Miras lie on a PL relation
as do those in the LMC, including the ones in LMC clusters discussed above.
LMC work (e.g. Whitelock et al. 2003) also shows that there are a small
number of large amplitude AGB variables which lie above the PL relation.
Providing such stars are also rare in our Galaxy they should not
significantly affect any of our conclusions. It has been suggested
(Whitelock \& Feast 2000, Whitelock et al. 2003) that these rare
overluminous stars are undergoing Hot Bottom Burning (HBB) - although this
is controversial in the case of C-rich stars as HBB is expected to convert
the carbon to nitrogen. Theory (e.g. Sackmann \& Boothroyd 1992) then
suggests that these stars probably have masses in excess of 4$M_{\odot}$,
significantly greater than those of normal Miras in the same period range.
It is therefore of interest that one such star discussed by Whitelock et
al., IRAS\,04496--6958, is possibly a member of the small cluster HS33. Van
Loon et al. (2005) obtained a colour-magnitude diagram of this cluster.
Whilst separation of the cluster from the field stars is difficult they
suggest an age in the range 110 to 200 Myr and hence a relatively massive
progenitor for IRAS\,04496--6958 in qualitative agreement with HBB
predictions.}

There is a suggestion from the velocity dispersions that at a given 
period the C-Miras are
younger than the O-Miras. The period-luminosity relation for C-Miras
(Whitelock et al. 2005) and that for O-Miras (Feast et al. 1989) 
suggest that also,
at a given period, the C-Miras are fainter 
bolometrically than the O-Miras,
although the thin-shelled members of these two classes
obey the same PL($K$) relation. Caution
is required in interpreting this since the spectral energy distribution
is different in the two types of Miras and may lead to different systematic
errors in deriving the bolometric magnitudes. 
For instance thin-shelled O-Miras whose energy distributions peak
in the near infrared may have estimated bolometric magnitudes which are
one or two tenths of a magnitude too bright because the $JHKL$ filters avoid
the main stellar water absorption bands (Robertson \& Feast 1981).
Whilst
accurate bolometric magnitudes are needed for comparison with theoretical
predictions, 
possible systematic errors of this type are not important in using
Miras as distance indicators as long as the bolometric luminosities
are calculated in a consistent manner. If despite the above caveat,
there is a real difference in the bolometric luminosities of
O- and C-Miras of the same age, this would be difficult to understand
unless, for instance, the two groups had different overall metallicities.
In that case, objects in one group (presumably that of lower metallicity) 
may have
become carbon stars before, or immediately on, 
entering the Mira stage. The work of
Abia et al. (2001) discussed at the end of our section 4, nevertheless,
suggests that carbon stars in the local galactic disc have metallicities
close to that of the disc mean, though a fuller understanding of the
metallicities of disc O- and C-Miras is very desirable.

It is striking that in the work on obscured Miras in the Magellanic Clouds
(Whitelock et al. 2003) there are no C-Miras with periods greater than 1000
days. The longest period C-Mira in our galactic sample is 815 days and there
is an object with a possible period of 1060 days in the sample of C-Miras
discussed by Groenewegen et al. (1998). However, the details of this star
have not, so far as we know, been published. We have therefore not included
it in the present discussion.  On the other hand the LMC sample of Whitelock
et al. contains a group of O-Miras with periods in the range 1097 to 1393
days. The absence of such long period O-Miras in the galactic Bulge, except
in the central region (Glass et al. 1995, Wood et al. 1998), suggests that
they are too young to be there.  They may well be younger than the longest
period C-Mira.  If that is the case they may be relatively massive objects
which have been prevented from becoming C-Miras by Hot Bottom Burning
(Boothroyd et al. 1993).

\section*{Acknowledgments}
We thank the referee, Jacco van Loon, for his comments.

\end{document}